\documentclass{aa}
\usepackage{astron}
\usepackage{graphics}
\usepackage{psfig}
\usepackage{times}

\def\etal{et al.}

\begin{document}
\thesaurus{03(11.01.1; 11.19.2; 11.19.3)}
\title{The formation of bars and disks in Markarian starburst galaxies}

\author{R. Coziol\inst{1}
       \and S. Consid\`ere \inst{1}
       \and E. Davoust \inst{2}
       \and T. Contini \inst{3}
}

\offprints{coziol@obs--besancon.fr}

\institute{Observatoire de Besan\c{c}on, UPRES--A 6091, B.P. 1615,
            F--25010 Besan\c{c}on Cedex, France
            \and
            Observatoire Midi--Pyr\'en\'ees, UMR 5572, 14 Avenue E. Belin,
            F--31400 Toulouse, France
            \and
            European Southern Observatory, Karl--Schwarzschild--Strasse 2,
             D--85748 Garching bei M\"unchen, Germany
}

\date{Received 20/09/1999; accepted 14/01/2000}

\authorrunning{Coziol et al.}
\titlerunning{The formation of bars and disks in Markarian galaxies}
\maketitle

\begin{abstract}

We have proposed in a companion paper (Consid\`ere et al. 2000)
that bars appeared recently in massive starburst nucleus galaxies.
We now test this hypothesis on an extended sample of barred and
unbarred Markarian starburst galaxies, using several samples of normal galaxies as
control samples.

In support of this hypothesis, we show that the
proportion of barred galaxies is much lower in
Markarian starburst galaxies than in normal galaxies.
In addition to this deficiency of bars, we find that
Markarian starburst galaxies have smaller disks than
normal galaxies, and that the disks
of unbarred starburst galaxies are smaller, on average,
than barred ones. Finally, we show that the Markarian starburst
galaxies do not seem to follow the local Tully--Fisher relation.

Various alternatives are examined to explain
the deficiency of bars and the small
disk dimensions in Markarian starburst galaxies.
One possibility, which is in agreement with the young
bar hypothesis, is that the formation of disks happens
after the formation of bulges and that bars appear only later,
when enough gas has been accreted in the disk.

\end{abstract}



\section{Introduction}

Since their discovery some thirty years ago,
starburst galaxies have been a swiftly growing subject of interest.

This interest is stimulated today by the discovery of many high-redshift
star forming galaxies which may have characteristics similar to the
nearby starburst galaxies (Steidel \etal\ 1996, 1999; Lilly \etal\ 1999).
However, the nature of the starburst phenomenon still eludes understanding.

When the massive starburst nucleus galaxies (SBNGs)
were discovered, for example, it was generally thought that they were
well-evolved galaxies which had been rejuvenated by
interactions with nearby companions \cite{Huchra77,TinsleyLarson79,Kennicutt90}.
But as many observations have shown,
starburst galaxies generally do not reside in high galaxy density regions
\cite{Iovinoetal98,COZIOLetal97a,Hashimotoetal98,Lovedayetal99},
which favor interactions, and only a fraction of them (between 25\% and 30\%)
have obvious luminous companions \cite{TellesTerlevich95,COZIOLetal97a}.

The assumption that SBNGs are well-evolved galaxies is also challenged by
observations. SBNGs generally have abnormal chemical abundances: they
are metal-poor compared to normal galaxies with
similar morphology and luminosity \cite{COZIOLetal97b} and they also
present an unusual excess of nitrogen abundance
in the nucleus\cite{COZIOLetal99}.

It appears that SBNGs are engaged in an important phase of formation
of their stellar population, but also of their chemical constituents. One
simple explanation of these phenomena is
that nearby SBNGs are examples of ``young'' galaxies
in their formation process. But is this the only alternative?

One popular assumption is that a bar
may be an efficient mechanism by which gas can accumulate
in the nucleus of an evolved galaxy to start a burst of star formation.
This structure would also produce some of the
chemical anomalies encountered in SBNGs. Indeed, it is generally
observed that normal barred spiral galaxies
have shallower metallicity gradients than unbarred ones
\cite{V-CE92,EdR93,ZARITSKYetal94,MR94}. From a theoretical point of view,
a bar is expected to funnel unprocessed gas
from its outer parts toward its nucleus
\cite{N88}, which decreases the metallicity gradient
by reducing the metallicity in the nucleus.

We have in fact verified
that bars cannot be at the origin of the nuclear bursts in SBNGs
(Consid\`ere et al. 2000).
We searched for the influence of the bar on star formation and chemical
evolution in a sample of 16 Markarian galaxies with strong bars and intense
star formation.  We studied the distribution
of ionized gas and the variations of oxygen  and
N/O abundance gradients along the bar.
No relations were found between these different
characteristics and the bar properties.

The aim of the present paper is to put the results of our
companion paper (Consid\`ere et al. 2000) in a broader context.
Using a large sample of galaxies, we test our interpretation that
bars in SBNGs appeared only
relatively recently. We show that this hypothesis is consistent with
a scenario where these galaxies are ``young'' galaxies still
in formation.

\section{The frequency of bars in Markarian starburst galaxies}

How can we further test whether bars in SBNGs are indeed young?
In Consid\`ere \etal\ (2000), we showed that
the bursts in the nuclei of SBNGs have not been triggered
by young bars, but by some event which probably took place
a few Gyrs in the past. In general, therefore,
the bursts in the nuclei of these galaxies
must be older than the bars.
We can use this assumed age difference to verify our hypothesis.
In a complete sample of SBNGs, one should expect
the frequency of barred galaxies to be proportional to the typical age of
bars divided by the typical age of nuclear bursts.
Therefore, if bars are young as compared to the nuclear
bursts, their frequency in the sample will be low.

To perform this test,
we have gathered information from the literature on all Markarian
galaxies (1500 galaxies), compiled by Mazzarella \& Balzano \cite*{MazzBalz86}.
Using LEDA, we have extracted 512 galaxies which had a morphological type,
were classified as starburst and were more luminous than magnitude M$_B = -18$.
We adopted this magnitude limit in order to select only SBNGs, and not HII
galaxies.  It is also known that the completeness of the Markarian sample
decreases for galaxy magnitudes M$_B \ge -18$ (Coziol
et al. 1997a). The present sub--sample represents one--third
of the whole sample of Markarian galaxies.
Although it cannot be considered statistically complete, it forms
the largest sub--sample of massive starburst galaxies
on which to apply our test\footnote{We remind
the reader that the starburst
galaxies studied by Consid\`ere \etal\ are all Markarian galaxies.}.

\begin{figure}
\resizebox{8cm}{!}{\includegraphics{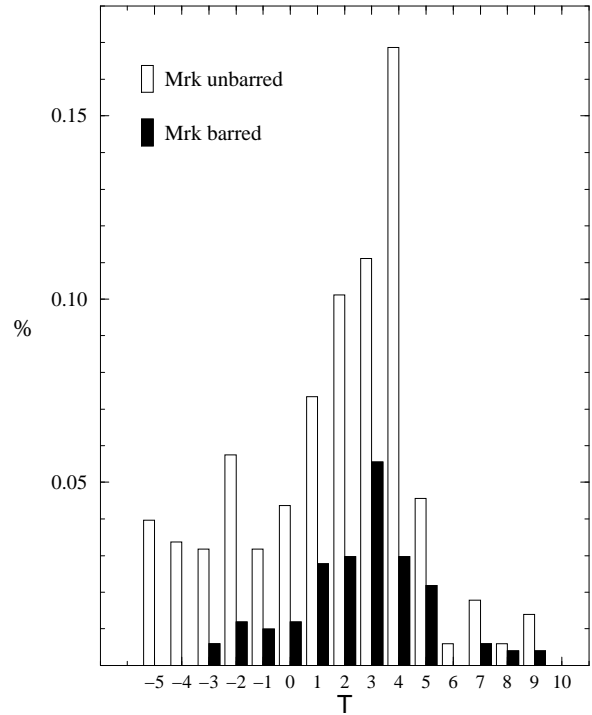}} \caption{
Distribution of morphologies in a sample of 512 Markarian galaxies
classified as starbursts. This sample represents one--third of the
Markarian catalogue. The histogram shows the percentage of the
total population in each morphological class. We also distinguish
between barred and unbarred galaxies }
  \label{Mark_T}
\end{figure}

The distribution of the morphologies of the Markarian starburst galaxies
is presented in Fig.~\ref{Mark_T}. We
find that only 109 Markarian starburst galaxies (21\%)
are barred. This frequency must be compared
to the frequency of bars in normal galaxies:
just over half of all normal galaxies are considered barred
\cite{dV63,SW93}.

The above statistics may obviously be biased, if many
unseen bars are present in the sample, as bars are often difficult
to detect in the optical (see for example Eskridge \etal 1999).
However, such a large discrepancy between normal and
Markarian starburst galaxies cannot be attributed to observational biases
only.
Even if we assume that many more Markarian starburst galaxies are
barred, we would need a proportion of barred galaxies significantly larger
than 50\% (the standard proportion in normal galaxies)
to confirm that bars play an important role
in the starburst phenomenon.

If the Markarian starburst sample were complete, the fraction of bars would
indicate exactly how old bars are in comparison to
the nuclear bursts. But, because of incompleteness,
the above frequency can only give
a qualitative estimate of the age difference.
In Consid\`ere \etal\ (2000) we estimated that
the nuclear bursts could be a few $10^9$\ yrs old, while
bars may be only a few $10^7$\ yrs old, which leads to an estimated
frequency of barred galaxies of 1\%.  This is lower than observed
by only a factor 10. Taken at face value, this
means that the bursts may be younger -- or the bars older -- than estimated
by a factor 10.
These two possibilities (or any solution in between) are consistent with our
observations. Taking into account the crudeness of our age estimates and
the conditions of the test, this result is reasonable.

\begin{table*}
\caption[]{Comparison of Markarian starburst and normal spiral
galaxy properties}
\begin{flushleft}
\begin{tabular}{lccccccc}
\noalign{\smallskip} \hline \hline \noalign{\smallskip} Sample &
\# gal. & $<M_B>$ & $<\mu_B>$           & $<Dist.>$ & $<R_{25}>$
&$med(R_{25})$ &$med(R_{25})$ (Sab/Sb) \\
       &         &       & (mag arcsec$^{-2}$) & (Mpc)     & (kpc)      &  (kpc)        &  (kpc) \\
\noalign{\smallskip} \hline \noalign{\smallskip} Normal unbarred &
437 & $-20.7\pm5.4$ & $23.5\pm0.5$ & $55\pm25$ & $31\pm13$& 26.9 &
30.6\\ Normal barred   & 598 & $-20.2\pm1.0$ & $23.5\pm0.5$ &
$44\pm23$ & $29\pm12$& 26.1 & 32.7\\ Mark unbarred   & 393 &
$-20.5\pm0.8$ & $22.8\pm0.8$ & $92\pm25$ & $20\pm10$& 17.7 &
19.0\\ Mark barred     & 109 & $-20.4\pm1.0$ & $23.0\pm0.6$ &
$67\pm23$ & $23\pm12$& 22.0 & 24.5\\ \noalign{\smallskip} \hline
\hline
\end{tabular}
\end{flushleft}
\label{TPARAM}
\end{table*}

From the above analysis, we conclude that these statistics support
the conclusions of Consid\`ere et al. (2000), that bars are not at
the origin of the nuclear bursts in SBNGs, because they are too
young and appeared only recently in these galaxies.

\section{The formation of the disk in Markarian starburst galaxies}

\begin{figure}
\resizebox{8cm}{!}{\includegraphics{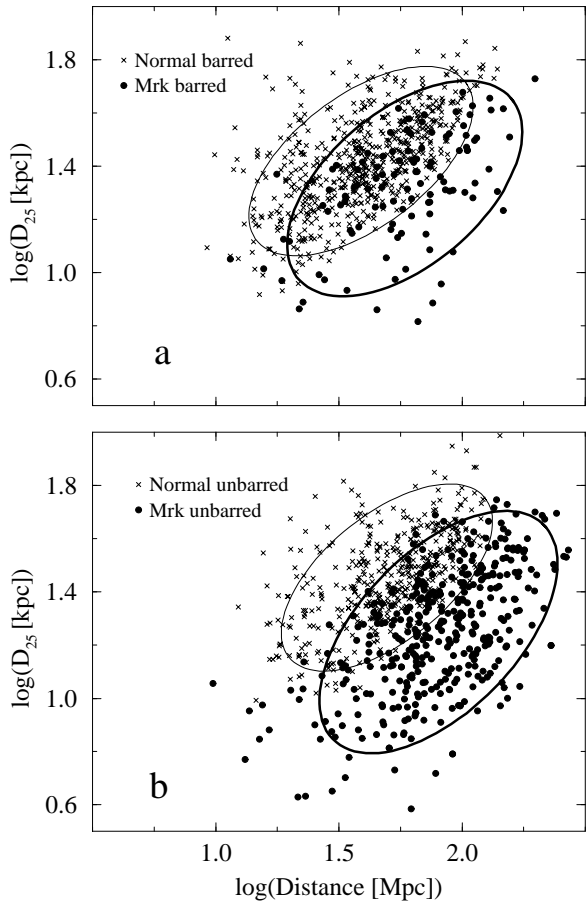}} \caption{ Disk
radius $R_{\rm 25}$, as a function of galaxy distance. {\bf a} :
the sample of normal barred spiral galaxies of Mathewson \& Ford
(1996) is compared to the Markarian barred starburst galaxies.
{\bf b} : the unbarred normal and Markarian spiral galaxies are
compared. The $2 \sigma$ dispersion ellipses are shown. The thin
ones encircle the normal galaxies and the thick ones the Markarian
starburst galaxies. In general, the Markarian starburst galaxies
have smaller disks than normal galaxies}
  \label{LogD_disk}
\end{figure}

The next step is to relate the recent appearance of bars in SBNGs
to the starburst phenomenon. In earlier papers, we showed that
SBNGs are probably the remnants of mergers of gas--rich and
small--mass galaxies, a process which we identified with
hierarchical formation (Coziol \etal\ 1997b;1998). We also found
an interesting difference between the chemical abundance of
early-- and late--type starburst galaxies suggesting that their
chemical evolution followed slightly different paths, namely that
late--type starburst galaxies accreted more gas than stars during
their formation \cite{COZIOLetal98}. This suggests an alternative
scenario for the disk formation in late--type starburst galaxies.

The merging of gas-rich and small-mass galaxies -- main origin of
the bursts -- produced the bulk of stars and chemical elements.
Depending on the density of the environment, this first phase of
formation produced galaxies with different bulge/disk ratios, with
a bias towards higher ratios, because mergers favor the formation
of early--type galaxies. But because the galaxy spatial density
where these galaxies formed is relatively low (starburst is a
field phenomenon), an important fraction of the gas did not
collapse and subsisted in a temporary reservoir around the
galaxies. As time passed, the reservoir emptied as the gas fell on
the galaxies, forming or increasing their disks. When the disk had
accreted enough gas a bar appeared.

The validity of the above scenario can be checked in the following
way. If starburst galaxies are still in the process of formation,
they should have smaller disks than normal galaxies. Moreover,
barred starburst galaxies should have larger disks on average than
unbarred ones. To test these predictions, we compare the isophotal
disk radii ($R_{\rm 25}$) of Markarian barred and unbarred
starburst galaxies with those of normal galaxies.  The result is
shown in Fig.~\ref{LogD_disk}, where the disk sizes are presented
as a function of distance. The normal barred and unbarred spiral
galaxies are represented by the sample of Mathewson \& Ford
\cite*{MathewsonFord96}. The disk radii were all extracted from
LEDA \cite{Patureletal97}. The mean radii are presented in
Table~\ref{TPARAM}. Unbarred starburst galaxies do seem to have a
smaller disk on average than barred ones. Moreover, starburst
galaxies generally have smaller disks than normal (barred and
unbarred) spiral galaxies.

\begin{figure}
\resizebox{9cm}{!}{\includegraphics{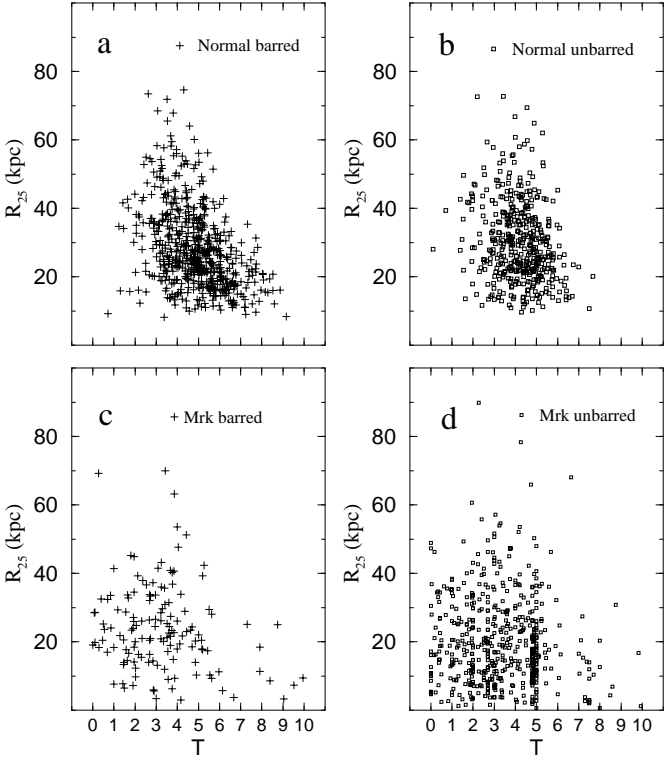}} \caption{Disk
radius $R_{\rm 25}$, as a function of galaxy morphology. {\bf a}
and {\bf b}, the normal barred and unbarred spiral galaxies of
Mathewson \& Ford (1996). {\bf c} and {\bf d}, the Markarian
barred and unbarred starburst galaxies. The small dimension of the
Markarian galaxies is independent of morphology }
  \label{LogDb_disk}
\end{figure}

It is important to understand the various biases which affect the two samples
of galaxies. Because of the Malmquist bias, the normal
galaxy sample is biased towards brighter (and thus larger)
galaxies as the distance increases.
This bias artificially raises the mean dimension of these
galaxies. But the Markarian sample is affected by the
same bias; even more so, because the sample goes slightly
deeper in redshift (see Table~\ref{TPARAM}).
The difference between normal spiral galaxies
and Markarian starburst galaxies, therefore, cannot
be explained by such a bias.
Moreover, the
difference observed between the disk radii of barred and unbarred
Markarian starburst galaxies cannot be attributed to the Malmquist
bias, since unbarred starburst galaxies
are on average located further away than
barred ones (see Table~1).

The above statistics may still be affected by another bias,
because the size of galaxies depends on their morphological type.
Roberts \& Haynes \cite*{RH94} have shown that the median
isophotal radius for a sample of 7930 galaxies in the UGC
catalogue varies with morphology. The radius is maximum for
intermediate types Sab/Sb and falls for earlier and later types.
How does this affect our result?

In Fig.~\ref{LogDb_disk}, we present the isophotal radius
of the Markarian starburst galaxies as a function of morphology.
As a comparison sample, we could not use that of Roberts \& Haynes and
used Mathewson \& Ford's sample of normal spiral galaxies
instead. One can see that
the size of the Markarian galaxies does not depend on
morphology. The barred and unbarred normal galaxies
share the same distribution in morphology, concentrated
around types Sb/Sbc. The Markarian galaxies, on the
other hand, are more numerous in the types Sab/Sb.
We thus have to be careful when comparing the different samples.

The incompleteness of Robert \& Haynes' sample is well
identified: the sample is deficient in low--surface brightness galaxies and
high surface brightness compact galaxies. These two deficiencies certainly do
not affect the comparison with our sample of Markarian galaxies, which are
neither compact nor of low surface brightness.

In their analysis, Robert \& Haynes used the median radius because their
distributions are not gaussian.
The median radius of their sample for Sab/Sb types is 25.1 kpc.
This has to be compared with the medians observed
in the samples of Markarian and Mathewson \& Ford (last
column in Table~1).
The medians for all the morphological types is also given in Table~1.
The spiral galaxies in the sample of Mathewson \& Ford
have slightly higher median values than
Roberts \& Haynes' sample. This is probably because
of the Malmquist bias. While the Markarian
barred galaxies are comparable in size to galaxies
in the sample of Robert \& Haynes, unbarred ones
are significantly smaller.

\section{The effect of small disks on the Tully--Fisher relation}

Do the small disks of Markarian
starburst galaxies affect their kinematics? In normal spiral
galaxies, the maximum rotation velocity is correlated to the absolute
magnitude by the Tully--Fisher (TF) relation. According to this
relation, massive galaxies have to rotate
more rapidly than small-mass galaxies in order to sustain their mass.
Table~1 shows that the Markarian galaxies have
luminosities and surface brightnesses which are comparable to those of normal
galaxies.  We thus expect them to be slow rotators if they follow the TF
relation, assuming that they have a normal mass-luminosity ratio
($\cal{M}/L$).

\begin{figure*}
\resizebox{18cm}{!}{\includegraphics{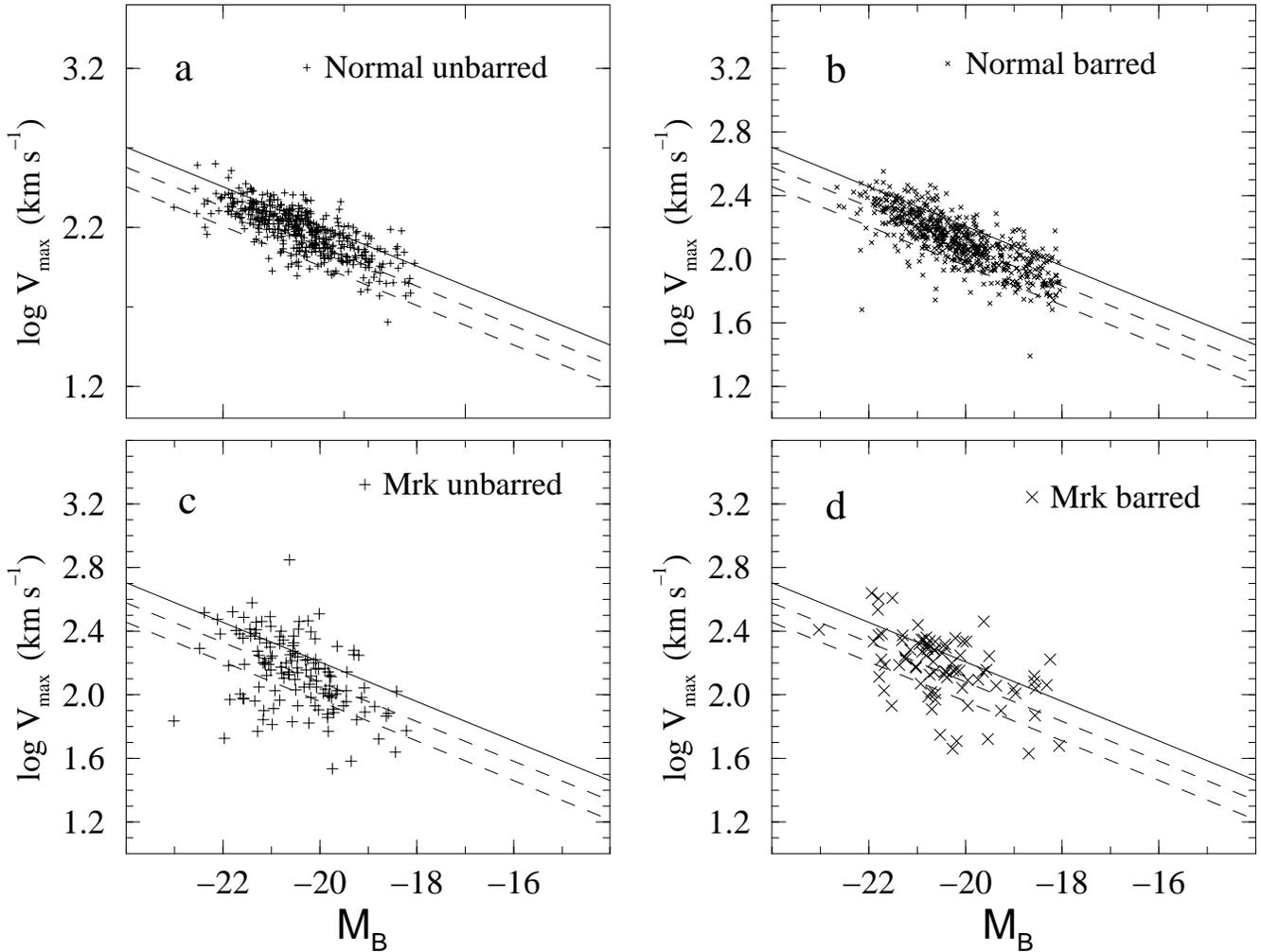}}
\caption{Tully--Fisher relation for ({\bf a}) normal unbarred and
({\bf b}) barred spiral galaxies; ({\bf c}) Markarian unbarred and
({\bf d}) Markarian barred starburst galaxies. The continuous line
is the local Tully--Fisher relation determined by Simard \&
Pritchet (1998). The dashed lines represent the same relation
shifted by $\Delta M_{\rm B} = 1.0$ mag and 2.0 mag}
 \label{Mark_TF}
\end{figure*}

We have determined the TF relation for the Markarian starburst
galaxies, using the maximum rotation velocities found in LEDA. In
Fig.~\ref{Mark_TF}, the starburst galaxies are compared to the
normal barred and unbarred spirals from Mathewson \& Ford (1996).
Using the latter sample, Simard \& Pritchet \cite*{SP98}
determined the local TF relation (the continuous line in
Fig.~\ref{Mark_TF}). The use of rotation velocities found in LEDA
gives a slightly lower value than the one found by Simard \&
Pritchet for the local TF relation, as the observed points tend to
fall slightly below the line.

We find that the Markarian starburst galaxies are not slow
rotators. They have rotation velocities comparable to those of
massive normal galaxies.  Then how do these galaxies readjust
their structure to follow the local TF relation ? In a galaxy in
dynamical equilibrium, the mass $\cal{M}$ is proportional to
$V_{max}^2 R$, where  $R$ is the radius of the disk and $V_{max}$
is the maximum rotational velocity. On the other hand, the total
luminosity $L$ is proportional to $\Sigma_0 R^2$, where $\Sigma_0$
is the central surface brightness.  Combining the two, we obtain :

\begin{equation}
R\ \Sigma_0\ ({\cal M}/L) \propto V_{max}^2
\end{equation}

In other words, a smaller galaxy must either have a higher surface
brightness or a higher ($\cal{M}/L$) to fall in the same region of the
TF relation as normal galaxies.  Since the Markarian starburst galaxies
have normal surface brightnesses (see Table~1), they should have
a higher ($\cal{M}/L$).  But this conclusion goes contrary
to what is usually admitted for starburst galaxies, where the presence
of massive stars raises the luminosity at the expense of the mass.

What appears as a contradiction might perhaps not be one, if one looks
more closely at Fig.~\ref{Mark_TF}.  The Markarian starburst galaxies
in fact do not follow any linear relation, although their rotational
velocities are consistent with values predicted by the local TF
relation. A linear regression applied to the unbarred Markarian galaxies
yields a coefficient of correlation of only 43\%. The
correlation is slightly better for barred starburst galaxies with 53\%.
The dispersion of the data is indeed significantly higher
in the Markarian sample than in the normal one. This higher dispersion
is not caused by spurious data; we have eliminated Markarian galaxies
with large uncertainties in $V_{max}$ ($ >10$\ km/s) in
Fig.~\ref{Mark_TF}. Therefore, this is an intrinsic characteristic
of the sample: the Markarian galaxies do not seem to
follow the local TF relation.

Although the origin of the TF relation
is still ill--understood, it is generally believed that
this relation is fundamental and that it
must have arisen at the formation of galaxies
\cite{Burstein1983,Bosch99,Steinmetz99,Syer99}.
The present observation is consistent
with this assumption, since we believe SBNGs are still
in their formation phase. It implies that the disks in these
galaxies are not in a state of dynamical equilibrium.

\section{Alternatives to the disk formation scenario}

Are there other ways to explain the paucity of barred galaxies in
the Markarian sample and their small disk dimensions?
A majority of Markarian galaxies might not have a bar
because the conditions required for the occurrence
of a burst do not allow it.
It is considered, for instance, that interactions can destroy the bar.
But in Consid\`ere et al. (2000) we have found
three clear cases of interacting galaxies where
the bars seem as strong as, if not stronger than other bars
in the galaxy sample.  Furthermore, this hypothesis does not explain the small
dimensions of the disks.

The fewer bars and the smaller disks in starburst galaxies could be due to
higher dust extinction. A high level of obscuration is effectively observed
in some ultraluminous infrared starburst galaxies \cite{MIRABELetal98}.
However, these objects are more an exception than the rule in the
nearby Universe. High extinction does not generally apply to
starburst galaxies \cite{BuatBurg98}. Moreover, in this
case the occurrence of smaller disks would imply that the dust opacity
becomes higher in the outer disk, while the contrary is usually found:
spiral galaxies are optically thin
in the outer regions and moderately opaque at their center
\cite{GIOVANELLIetal94,MORIONDOetal98,XILOURISetal99}.

There is also clear
evidence that the outer regions of disks are relatively unevolved at
the present epoch \cite{FERGUSONetal98}, which is consistent with the idea
that disks are younger than bulges, as proposed in our scenario.

As a last alternative, it may be that, at the end of their evolution, the
Markarian galaxies will produce mostly small disks and unbarred galaxies.
A similar hypothesis was recently suggested to explain the appearance of a
large number of
relatively small galaxies with high luminosities at high redshifts
\cite{LILLYetal98}.  If what we observe in nearby starburst galaxies is of
the same nature, then this means that at each epoch we always see some intense
star forming activity which concerns only galaxies of small dimensions.  This
is an intriguing hypothesis which would imply that Markarian galaxies are of a
peculiar nature.

None of the above scenarios predicts that Markarian starburst galaxies
should not follow the local TF relation.
The dispersion observed in Fig.~\ref{Mark_TF}
for the Markarian starburst galaxies cannot be explained
assuming only higher dust extinction. This
hypothesis might work for galaxies which are above
the local TF relation in Fig.~\ref{Mark_TF}, but not for galaxies
which are below; they would need to be more luminous than normal.
It is also hard to understand why dust extinction
changes neither the distribution
of luminosity nor the surface brightness of these galaxies (see Table~1).
We need a very contrived model for dust distibution in order to
explain all these observations.

In their paper on star forming galaxies at high
redshifts, Simard \& Pritchet \cite*{SP98} found that these
galaxies do not follow the local TF relation. They concluded that this could
be explained by assuming that high redshift star forming
galaxies are more luminous, by an average of one or two magnitudes, than normal
nearby galaxies. But there is no evidence that
the B luminosity of the local Markarian starburst galaxies, and of SBNGs
in general, differs significantly from that of normal galaxies (see Table~1
and Coziol 1996). Furthermore, Simard \& Pritchet do not know if the
disks of their galaxies are smaller than those of normal galaxies,
as in nearby starburst galaxies.
If the Markarian starburst galaxies do not follow the local TF relation
because they are more luminous than normal spiral galaxies,
then, taking into account their small dimension,
nearby starburst galaxies must be much
more luminous than ``comparable'' star forming
galaxies at high redshifts. This argument suggests that
the reason why Markarian galaxies do not follow the local
TF relation is that they are still in the process of forming their
disks. This may be true also for forming galaxies at high redshifts.

\section{Conclusion}

Of all the alternatives presented above, none is simpler than our scenario for
the formation of the nearby SBNGs.  It has many
advantages:  it explains the origin of the bursts in these galaxies (the
galaxies are in a burst phase because they are still forming), and it
fits their star formation history \cite{COZIOL96}, their chemical
evolution (Coziol \etal\ 1998; 1999) and the properties of
their bars (Consid\`ere et al. 2000).

According to this scenario, bulges of galaxies form first and the disks form
later mostly through gas accretion.
It predicts that young galaxies initially look like
unbarred early--type spirals with small disks. Then, as the disk grows, they
change into late-type and giant barred spiral galaxies
\cite{KWG93,BCF96,ANDREDAKIS98}. This transformation may explain why
Markarian starburst galaxies are so frequent among Sa and Sb galaxies
(see Fig.~\ref{Mark_T}). The
fact that we do not see many Sc galaxies in the sample of Markarian
starburst galaxies suggests that these galaxies forms
differently \cite{ANDREDAKIS98}.

\begin{acknowledgements}
R. C. would like to thank the Observatoire de Besan\c{c}on and
Universit\'e de Franche-Comt\'e for
funding his visit, during which this paper was written. He would also like
to thank the direction and staff of Observatoire de Besan\c{c}on for their
hospitality. We acknowledge with thanks positive comments and constructive
suggestions from the referee, Danielle Alloin, which helped to improve
the quality of this paper.
For this research, we made use of the LEDA database.

\end{acknowledgements}


\end{document}